\documentclass[conference]{IEEEtran}
\IEEEoverridecommandlockouts
\usepackage{amsmath,amssymb,amsfonts}
\usepackage{algorithmic}
\usepackage{graphicx}
\usepackage{textcomp}
\usepackage{xcolor}
\usepackage[numbers,sort&compress]{natbib}
\def\BibTeX{{\rm B\kern-.05em{\sc i\kern-.025em b}\kern-.08em
    T\kern-.1667em\lower.7ex\hbox{E}\kern-.125emX}}
\begin{document}

\title{Fast Scenario Reduction for Power Systems by Deep Learning\\
}

\author{\IEEEauthorblockN{Qiao Li}
\IEEEauthorblockA{\textit{dept. of Electrical \& Computer Engineering} \\
\textit{University of Denver}\\
Denver, USA \\
lq0607@gmail.com}
\and
\IEEEauthorblockN{David Wenzhong Gao}
\IEEEauthorblockA{\textit{dept. of Electrical \& Computer Engineering} \\
\textit{University of Denver}\\
Denver, USA \\
wenzhong.gao@du.edu}
}

\maketitle

\begin{abstract}
Scenario reduction is an important topic in stochastic programming problems. Due to the random behavior of load and renewable energy, stochastic programming becomes a useful technique to optimize power systems. Thus, scenario reduction gets more attentions in recent years. Many scenario reduction methods have been proposed to reduce the scenario set in a fast speed. However, the speed of scenario reduction is still very slow, in which it takes at least several seconds to several minutes to finish the reduction. This limitation of speed prevents stochastic programming to be implemented in real-time optimal control problems. In this paper, a fast scenario reduction method based on deep learning is proposed to solve this problem. Inspired by the deep learning based image process, recognition and generation methods, the scenario data are transformed into a 2D image-like data and then to be fed into a deep convolutional neural network (DCNN). The output of the DCNN will be an "image" of the reduced scenario set. Since images can be processed in a very high speed by neural networks, the scenario reduction by neural network can also be very fast. The results of the simulation show that the scenario reduction with the proposed DCNN method can be completed in very high speed.
\end{abstract}


\section{Introduction}
With the rapid development of renewable energy in recent years, the penetration of renewable energy in power systems increases greatly. Also, the concept of microgrid is getting popular in industrial. A typical microgrid contains a big portion of renewable energy in its total generation. As a result, the randomness introduced by the renewable sources can influence the stability and efficiency of the power systems and microgrids greatly. On the other hand, the load in most power systems are also very random. Although many load forecasting methods are proposed to study the behavior of the load. It is still impossible to predict the load with no errors. Therefore, the control and optimization based on the forecasted data may not be the best way to operate the power system.

To solve this problem, the stochastic programming is adopted in many works to deal with the randomness and stochastic phenomenon of renewable energy and load. Stochastic programming can optimize the power systems according to different possible scenarios. In comparison, the traditional optimization methods only consider one scenario. By incorporating with stochastic programming, the better results of unit commitment and model predictive control (MPC) can be achieved \cite{necoara2014mpc, parisio2013stochastic, takriti1996stochastic, ozturk2004solution}. In order to perform stochastic programming, the scenario set for the power system is needed. The scenario set contains many different scenarios and each scenario represents one possible trajectory of the states of the power system. However, since a power system is a big system with a lot of states and each state can have many different possible values, the total number of scenarios in a scenario set is usually very big. It needs massive computation resources to perform the stochastic programming with the entire scenario set, which is not efficient. So, the scenario reduction method is invented to reduce the number of scenarios in the scenario set, thus the amount of data need to be processed is smaller. Several popular scenario reduction approaches are proposed. Backward and forward methods \cite{growe2003scenario} are the most known scenario reduction methods. These methods delete the redundant scenarios and keep the space distance between the original scenario set and the reduced scenario set small. However, backward and forward methods are very slow compare to other methods. To improve the speed of the forward method, the simultaneous backward reduction (SBR) and fast forward selection (FFS) methods is proposed in \cite{heitsch2003scenario}. But it is still not fast enough for big scenario set. In addition to backward and forward methods, some papers use PSO (Particle swarm optimization) to obtain the optimal reduced scenario set \cite{pappala2009stochastic, xiong2017pso}. PSO method has similar performance as the FFS method according to \cite{lin2017scenario}. Paper \cite{li2016scenario} introduces a heuristic search (HS) method to obtain the optimal reduced scenario set by considering both the space distance and moment distances to the original scenario set. In \cite{lin2017scenario}, the authors introduce an improved initial-center refined and weighted K-means (ICRW K-means) method for scenario reduction. This method improves the scenario reduction speed compare to PSO and FFS methods. However, the speed is still not enough for realtime control applications. According literature review, there is no deep learning-based method being used in scenario reduction. But there some works which implement the deep learning or neural network techniques in scenario generation problems. Paper \cite{vagropoulos2015artificial} uses ANN for one step-ahead load forecasting and then generate the scenario set of load by Gaussian white noise. A GAN (generative adversarial networks) model is employed in \cite{chen2018model} to learn the scenario generation, and then the GAN model is used for scenario generation for wind and solar power.

In this paper, a new scenario reduction method based on deep learning is proposed. Inspired by the deep learning-based image process, recognition and generation methods, which are usually very fast. A deep convolutional neural network (DCNN) model is developed for the fast scenario reduction of renewable system. In order to achieve the scenario reduction with the DCNN, the scenario set is transformed into a 2-D image-like matrix. So the DCNN can process the scenario set as the images. Then, the output of the DCNN will be an smaller "image" containing the data of the reduced scenario set. It should be note that this method does not actually reduce the original scenario set, but instead generate a smaller scenario set which is similar to the optimal reduced scenario set.

The major contributions of this paper are summarized as follows:
\begin{itemize}
\item A fast scenario reduction method is proposed. The proposed scenario reduction method can be fast enough for the realtime applications such as model predictive control.
\item To the best of our knowledge, this is the first deep learning-based method for scenario reduction problem. The way to treat the scenario set as an image in this method will be useful to implement other deep learning mdoels in scenario reduction problems.
\end{itemize}

This paper is organized as follows: Section II describes in detail the proposed scenario reduction methodology. The method to pre-process the scenario set data and the structure of the DCNN model will be introduced in different subsections. Section III presents the simulation results of the proposed method on solar power. Finally, conclusions are drawn in Section IV.

\section{Proposed Approach}

\begin{figure*}[htbp]
    \centerline{\includegraphics[width=\hsize]{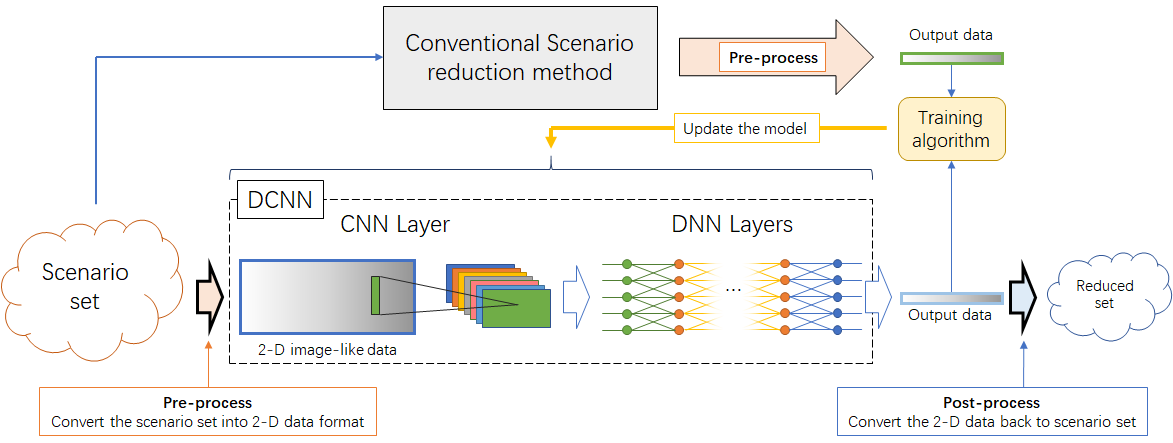}}
    \caption{The structure of the DCNN model.}
    \label{structure_all}
\end{figure*}

The proposed DCNN for scenario reduction is shown in Fig.\ref{structure_all}. The model reduces the scenario set $P\{\zeta_{s,1,...,T}, p_s\}$ with $S$ scenarios into a reduced scenario set $\hat{P}\{\hat{\zeta}_{s,1,...,T}, p_s\}$ with $\hat{S}$ scenarios. In this model, the original scenario set should be pre-processed into a 2-D data format at the first. Then the DCNN model will treat this 2-D data as an image and output a smaller image according to the input. The small image from the output of the DCNN is the reduced scenario set. In order to get a good reduced scenario set, the DCNN model is trained to generate the output similar to the reduced scenario set from other convectional scenario reduction method. The details about the DCNN model and data processing will be introduced in the following subsections.

\subsection{Data pre-process}
The DCNN model in this paper contains a CNN layer, which is usually used for image processing. In order to use the DCNN model the process the scenario set, the data of scenario set should be transformed into 2-D matrix format in which is same as an image. Suppose there is a scenario set $P\{\zeta_{s,1,...,T}, p_s\}$ of the total solar generation of a solar array in $T$ time steps. The $s$th scenario $\zeta_{s,1,...,T}$ in the set has probability $p_s$. There are $S$ scenarios in the set in total, so $s=1, 2, ..., S$. For this scenario set, the data pre-process module will transform the data of the scenario set into the following matrix,

\begin{equation}\label{2D_data}
\left[\begin{matrix}
      \zeta_{1,1} & \zeta_{2,1} & \cdots & \zeta_{S,1}\\
      \zeta_{1,2} & \zeta_{2,2} & \cdots & \zeta_{S,2}\\
      \vdots      & \vdots      & \ddots & \vdots \\
      \zeta_{1,T} & \zeta_{2,T} & \cdots & \zeta_{S,T}\\
      p_{1}       & p_{2}       & \cdots & p_{S}
\end{matrix}\right]
\end{equation}

where $\zeta_{s,t}$ for $s=1, 2, ..., S$ and $t=1, 2, ..., T$ are the states (solar generation) of scenario $s$ at time step $t$.
In this way, the scenario set is converted into a 2-D data format, i.e., an image with $(T+1) \times S$ resolution.

Since the scenario reduction only reduce the number of scenarios but not the length of each scenario, so the output of the DCNN should be a vector of $(T+1)\hat{S}$ elements (the output of a neural network is a vector), which represents a $(T+1) \times \hat{S}$ matrix. According to the Fig.\ref{structure_all}, the output data of the DCNN should be post-processed back to the scenario set data format. This conversion is same as the pre-process shown above but in the opposite direction.

\subsection{DCNN model}
As shown in Fig.\ref{structure_all}, the DCNN model includes two major parts: CNN layer and DNN (Deep Neural Network) layers.
The one CNN layer is used in the front of the model to extract the patterns or features in the original data. One the other hand, with the CNN layer, the data is compressed while the key features in the original data can be retained. This compression reduces the total amount of data the model need to deal with, hence less computation resources is needed. Inside the CNN layer, there is one convolutional layer and one pooling layer.
The convolution layer can filter out the key features and patterns in the data. The filters in the convolutional layer in this model is of $(T+1) \times w$ size. The number $w$ can be chosen with a small integer. With this filter shape, the filter can contain several complete scenarios. In this way, the features of the scenarios along the entire time period can be recognized.
In the pooling layer, the pool size is $1\times S/\hat{S}$. With this pool size, the input data will be reduced into the desired size by combining the nearby scenarios. Note that the size of the reduced scenario set $\hat{S}$ should be a factor of the size of the original scenario set $S$.
After the CNN layer, the data is input into the DNN model. The DNN model is a series of ANNs (Artificial Neural Network). In the DNN models, the number of filtered data generated by the CNN layer will decrease and eventually reach $1$, which is the desired output.
The detailed structure of each layer in the DCNN is presented in Table \ref{DCNN_stru_tab1}.

\begin{table}[htbp]
\caption{The DCNN model structure.}
\begin{center}
\begin{tabular}{|c|c|c|c|}
\hline
  & layer type & output size & activation\\
\hline
layer 1 & Convolutional layer & $(T+1) \times S \times 64$ & ReLU\\
layer 2 & Pooling layer & $(T+1) \times \hat{S} \times 64$ & ReLU\\
layer 3 & ANN & $(T+1) \times \hat{S} \times 32$ & ReLU\\
layer 4 & ANN & $(T+1) \times \hat{S} \times 8$ & ReLU\\
layer 5 & ANN & $(T+1) \times \hat{S}$ & Sigmoid \\
\hline
\end{tabular}
\label{DCNN_stru_tab1}
\end{center}
\end{table}

\subsection{Training}
The objective of this model is to generate the reduced scenario data similar to the reduced set obtained from other good scenario reduction method. In this paper, the heuristic search (HS) method from \cite{li2016scenario} is used to train the DCNN model. So, as shown in Fig.\ref{structure_all}, the HS method and the propsed DCNN model will be input with same scenario set, and then the output of the DCNN model will be compared with the output of the HS method to adjust the parameters of the DCNN model. In this paper, the binary cross-entropy is used to measure the difference between the DCNN output and the result of HS method. The AdaDelta optimizer is used as the training algorithm to update the parameters of the DCNN model.

\section{Simulation results}
\begin{figure}[htbp]
    \centerline{\includegraphics[width=\hsize]{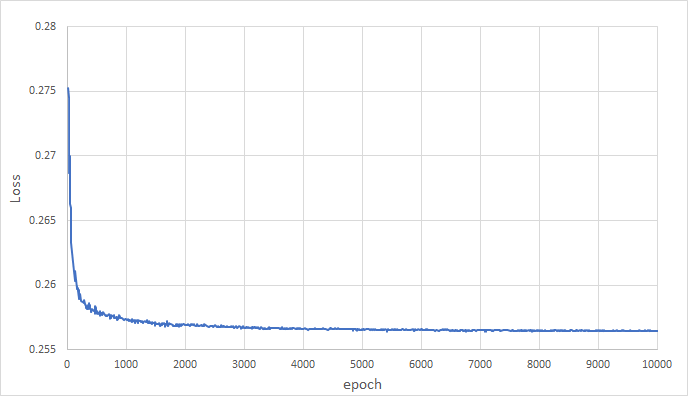}}
    \caption{The losses of the DCNN model during training.}
    \label{losses}
\end{figure}

In order to test the proposed DCNN model, the simulation of using the DCNN to reduce the scenario set of a 4kW solar array is provided. The data is obtained from PVWatts toolbox \cite{PVwatt_web} developed by NREL (National Renewable Energy Laboratory). The scenarios in the set represent the possible solar generation in 24 hours (one data point for each hour). The objective is to reduce the scenario set with $1000$ scenarios into the reduced scenario set with $200$ sceneries. The simulation is performed on a PC with Intel i7 2.8Ghz CPU, 24GB memory and NVIDIA GTX 1050Ti graphic card with 4GB graphic memory. The model is coded in Python with Tensorflow and accelerated by the GPU.
To train the DCNN model, $82\%$ data are used to train the model and the rest $18\%$ data are used to test the model. The model is trained by $10000$ epoches. The losses of the output during the training is shown in Fig.\ref{losses}.

\begin{figure}[htbp]
    \centerline{\includegraphics[width=\hsize]{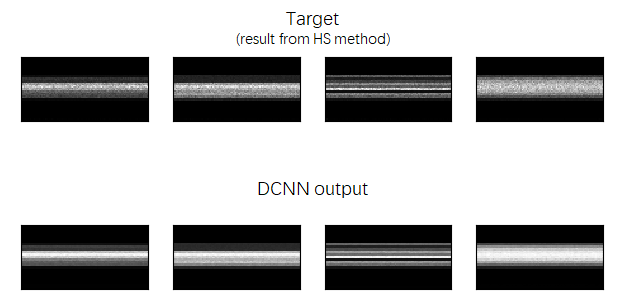}}
    \caption{The results of DCNN model and the HS model in image format.}
    \label{result_image}
\end{figure}

\begin{figure}[htbp]
    \centerline{\includegraphics[width=\hsize]{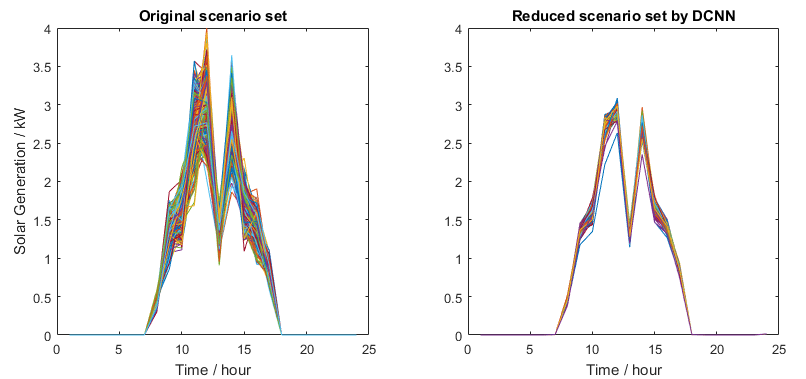}}
    \caption{The original scenario set vs the reduced set by DCNN.}
    \label{result_curves}
\end{figure}

Fig.\ref{result_image} shows some results of the DCNN model and the HS model (target) in image format. It shows that the output of the DCNN can mimic the target data, which is the reduced scenario set from HS method. While Fig.\ref{result_curves} shows one result of the reduced set compared with the original scenario set. The result shows that the reduced scenario set can represent the trend of the data in the original set.

\begin{table}[htbp]
\caption{Performance and accuracy of the DCNN model vs HS method.}
\begin{center}
\begin{tabular}{|c|c|c|}
\hline
& DCNN & HS method\\
\hline
Time consumed (s) &  0.0214 & 191.56\\
Space distance & 0.0056 & 0.0228\\
Moment distance & 1.43e-06 & 2.29e-07\\
\hline
\end{tabular}
\label{DCNN_HS_comp_tab1}
\end{center}
\end{table}

The performance and accuracy of the DCNN model is compared with the HS method in Table \ref{DCNN_HS_comp_tab1}. The results show that the proposed DCNN method can obtain a very good result as the HS method but with much fast speed.

\section{Conclusion}
This paper proposed a DCNN-based scenario reduction method. It is the first work of using deep learning method to deal with scenario reduction problem. The DCNN treats the scenario set as image, and thus be able to process the reduction in a very high speed.  To evaluate the prosed method, the DCNN model is simulated to reduce the scenario set of solar generation data. The results are compared with HS method. The results show that the proposed method can generate a good reduced scenario set within a very short time. With the high speed scenario reduction ability of this method, the stochastic programming can be implemented in realtime application in the future.

\bibliographystyle{IEEEtran}
\bibliography{ref}

\end{document}